\def\kms{\hbox{$~$km$~$s$^{-1}$}}
\def\snid{\ifmmode{\rm \sc SNID}\else{\sc SNID}\fi}

\def\catwo{Ca~{\sc ii}}

\def\fethree{Fe~{\sc iii}}

\def\stwo{S~{\sc ii}}
\def\sitwo{Si~{\sc ii}}

\documentclass[
    ,final            
  ]
  {aipproc}

\usepackage{amssymb}

\layoutstyle{6x9}

\begin{document}

\title{Determining the Type, Redshift, and Phase of a Supernova Spectrum}


\author{St{\'e}phane Blondin}{
  address={Harvard-Smithsonian Center for Astrophysics, Cambridge, MA
  02138}
}

\author{John L. Tonry}{
  address={Institute for Astronomy, University of Hawaii, Honolulu, HI
  96822}
}

\begin{abstract}
We present an algorithm to identify the types of supernova spectra,
and determine their redshift and phase. This algorithm, based on the 
correlation techniques of Tonry \& Davis, is implemented in the
SuperNova IDentification code (SNID). It is used by members of the
ESSENCE project to determine whether a noisy spectrum of a
high-redshift supernova is indeed of type Ia, as opposed to, e.g.,
type Ib/c. Furthermore, by comparing the correlation redshifts
obtained using SNID with those determined from narrow lines in the
supernova host galaxy spectrum, we show that accurate redshifts (with
a typical error $\sigma_z \lesssim 0.01$) can be determined for SNe~Ia 
for which a spectrum of the host galaxy is unavailable. Last, the
phase of an input spectrum is determined with a typical accuracy of
$\sigma_t \lesssim 3$ days. 
\end{abstract}

\maketitle


\section{Introduction}

Supernovae play a major role in the recent revival of observational
cosmology. It is through observations of type-Ia 
supernovae (SNe~Ia) over a large redshift range that two teams were
able to independently confirm the present accelerated rate of the
universal expansion \citep{Riess/etal:1998,Perlmutter/etal:1999}. This
astonishing result has been confirmed in subsequent years at moderate
redshifts \citep{Tonry/etal:2003,Knop/etal:2003,Barris/etal:2004}, but
also at higher ($z > 1$) redshifts where the universal expansion is in
a decelerating phase \citep{Riess/etal:2004}. Currently, two ongoing
projects have the more ambitious goal to measure the equation-of-state
parameter of the ``dark energy'' that drives the expansion: the
ESSENCE \citep{Wood-Vasey/etal:2007} and SNLS \citep{Astier/etal:2006}
projects. The success of these cosmological experiments depends,
amongst other things, on the assurance that the supernovae in the
sample are of the correct type, namely, SNe~Ia. Inclusion of
supernovae that are of a different type or exclusion of SNe~Ia from
the sample leads to biased cosmological parameters in the former case
\citep{Homeier:2005} and increased statistical errors on these same
parameters in the latter. The secure classification of supernovae
is a challenge at all redshifts, however. Even with high
signal-to-noise ratio (S/N) spectra, the distinction between supernova
of different types (or between subtypes withing a given type) can pose
problems.

The spectrum of a supernova also contains information on its redshift
and phase. Knowledge of the SN redshift is necessary for the use of
SNe~Ia as distance indicators (though see \citep{Barris/Tonry:2004}
for redshift-independent distances), and is usually determined {\it
via} narrow lines in the spectrum of the host galaxy. When such a
spectrum is unavailable, however, one has to rely on comparison with
SN template spectra for determining the redshift. The SN phase is
usually determined using a well-sampled lightcurve, but a single
spectrum can also provide a relatively accurate estimate. Moreover,
comparison of spectral and lightcurve phases of high-redshift
supernovae can be used to test the general relativistic prediction of
time dilation \citep{Riess/etal:1997,Foley/etal:2005}.

We have developed a tool (SuperNova IDentification; SNID) to determine
the type, redshift and phase of a supernova, using a single
spectrum. The algorithm is based on the correlation techniques of 
Tonry \& Davis \citep{Tonry/Davis:1979}, and relies on the comparison
of an input spectrum with a database of high-S/N template spectra. Our
database presently comprises 796 spectra of 64 SNe~Ia, 172 spectra of
8 SNe~Ib, 116 spectra of 9 SNe~Ic, 353 spectra of 10 SNe~II, as well
as spectra of galaxies, AGNs, and variable stars. The supernova
spectra cover a broad range of phases, and span a sufficient restframe
wavelength range ($\lambda_{\rm min} \le 4000$\,\AA; $\lambda_{\rm max} \ge
6500$\,\AA) to include all the identifying features of SN
spectra. Most of these spectra are publicly available through the
SUSPECT\footnote{\url{http://bruford.nhn.ou.edu/~suspect/index1.html}}
and
CfA\footnote{\url{http://www.cfa.harvard.edu/oir/Research/supernova/SNarchive.html}}
SN spectral archives.

We briefly describe the cross-correlation technique in the next
section, and then test the accuracy of correlation redshifts and
phases using SNID. Last, we tackle the issue of supernova
classification by focusing on two examples relevant to SN searches at
high redshifts.


\section{Cross-correlation techniques}

The cross-correlation method presented here is extensively discussed
by \citet{Tonry/Davis:1979}, where it is exclusively applied to galaxy
spectra. Nonetheless, the formalism is easy to adapt to supernova
spectra. The correlation technique is in principle fairly
straightforward: a supernova spectrum, $s(n)$, whose redshift, $z$, is
to be found is cross-correlated with a template spectrum, $t(n)$ (of
known type and phase) at zero redshift. We want to determine the
$(1+z)$ wavelength scaling, that maximizes the cross-correlation $c(n)
= s(n) \star t(n)$. In practice, it is convenient to bin the spectra
linearly with $\ln \lambda$, where $\lambda$ denotes the
wavelength. Scaling the wavelength axis of $t(n)$ by a factor $(1+z)$
is then equivalent to adding a $\ln (1+z)$ shift to the logarithmic
wavelength axis of  $t(n)$, i.e. a (velocity) redshift corresponds to
a uniform linear shift.

\begin{figure}
\includegraphics[height=.67\textwidth]{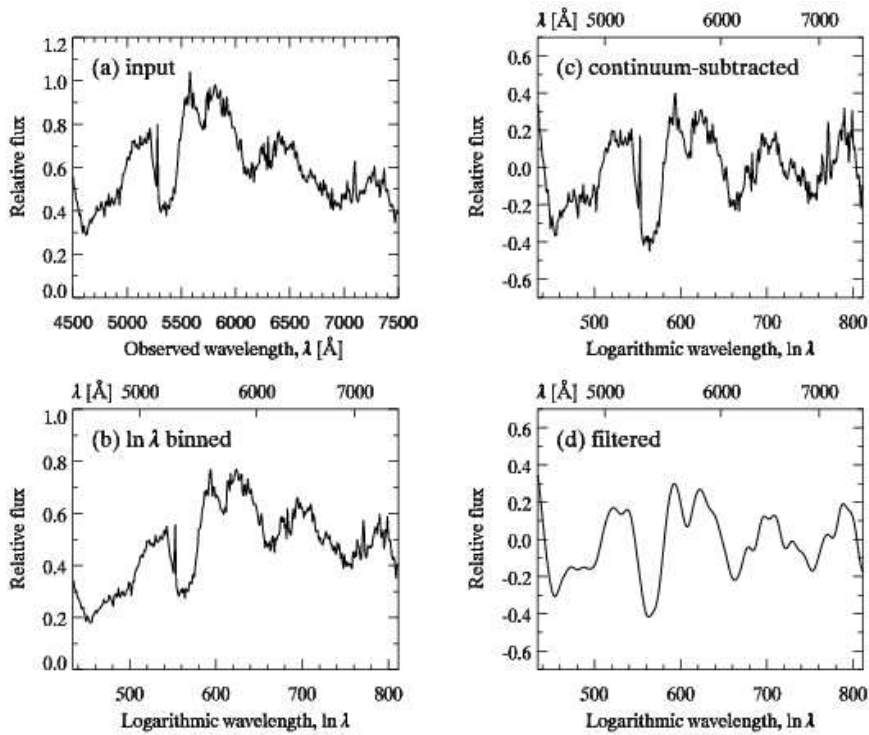}
\caption{Spectrum pre-processing. The input spectrum (a) is binned on
a logarithmic wavelength scale (b). A {\it pseudo}-continuum is
subtracted (c) and the resulting spectrum is bandpass filtered (d).
\label{Fig:lnlambda}}
\end{figure}

We show the result of mapping an input supernova spectrum onto a
logarithmic wavelength axis in Fig.~\ref{Fig:lnlambda}. We show the
input spectrum in panel (a) and its $\ln \lambda$ binned version in
panel (b). The next step in preparing the spectra for correlation
analysis is {\it pseudo}-continuum subtraction
(\citep{Tonry/Davis:1979}; Fig.~\ref{Fig:lnlambda}, panel (c)). The
purpose is to effectively remove any intrinsic color information in
the input and template spectra, and to ensure the correlation is not
affected by reddening uncertainties or instrumental distortions. This
effectively discards any spectral color information, and the
correlation only relies on the {\it relative} shape and strength of
spectral features in the input and template spectra. The final step is
the application of a bandpass filter (Fig.~\ref{Fig:lnlambda}, panel
(d)). The goal is to remove low-frequency residuals left over from the
pseudo-continuum subtraction and high-frequency noise components.

\begin{figure}
\includegraphics[height=.45\textwidth]{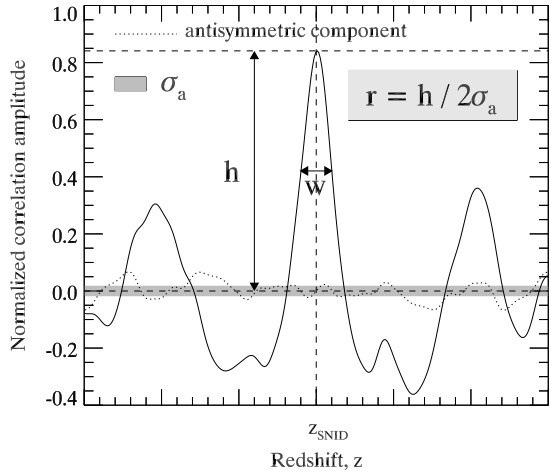}
\caption{The correlation $r$-value is defined as the ratio of the
height, $h$, of the highest peak in the normalized correlation
function ({\it solid line}) to twice the RMS, $\sigma_a$, of its
antisymmetric component ({\it dotted line}).
\label{Fig:rdef}}
\end{figure}

\citet{Tonry/Davis:1979} introduce a parameter, $r$, to quantify the
significance of a correlation peak in $c(n)$. It is defined as the
ratio of the height $h$ of the peak to the root-mean-square (RMS),
$\sigma_a$ , of the antisymmetric component of $c(n)$ about the
correlation redshift (see Fig.~\ref{Fig:rdef}). A perfect correlation 
will have a peak with $h=1$ at the exact redshift, and $c(n)$ will be
symmetric about this redshift, i.e. $\sigma_a=0$ and $r
\longrightarrow \infty$. Conversely, $r$ will be small ($r < 3$) for a
spurious correlation peak, and large ($r \gtrsim 8$) for a significant
peak, since $h$ will be close to 1 and $\sigma_a$ will be small. The
width of the correlation peak, $w$, is used to formally evaluate the
redshift error (not discussed here).

The $r$-value is further weighted by the overlap in $\ln \lambda$
space (at the correlation redshift) between the input spectrum and
each of the template spectra used in the correlation. The template
spectra are trimmed to match the wavelength range of the input
spectrum at the redshift corresponding to the correlation peak. The
overlap value, $lap$, conveys important absolute information about the
quality of the correlation, complementary to the correlation parameter
$r$. We usually discard correlation redshifts that have an associated
$lap < lap_{\rm min} = 0.40$ and a combined $rlap = r \times lap <
rlap_{\rm min} = 5$.


\section{Redshift and Phase Determination}

We use a simple simulation to test the accuracy of \snid\ in
determining the redshift and phase of a supernova spectrum. Here we
only consider ``normal'' type-Ia supernovae since they are the most 
represented in our spectral database, though the conclusions announced
in this section are qualitatively valid for all other supernova types.

In this simulation, each normal SN~Ia spectrum in the database is
correlated with all other normal SN~Ia spectrum using \snid, after
having taken care to temporarily remove all spectra corresponding to
the input supernova from the database. The input spectrum is first
redshifted at random in the interval $0.1 \le z \le 0.7$. We then add
noise (both random Poisson noise and sky background) to reproduce the
range of typical signal-to-noise ratio of SN spectra at the simulation
redshifts, when observed with 8-10\,m-class telescopes (e.g., VLT,
Keck, Gemini) used in cosmological SN~Ia surveys.

\begin{figure}
\includegraphics[height=.45\textwidth]{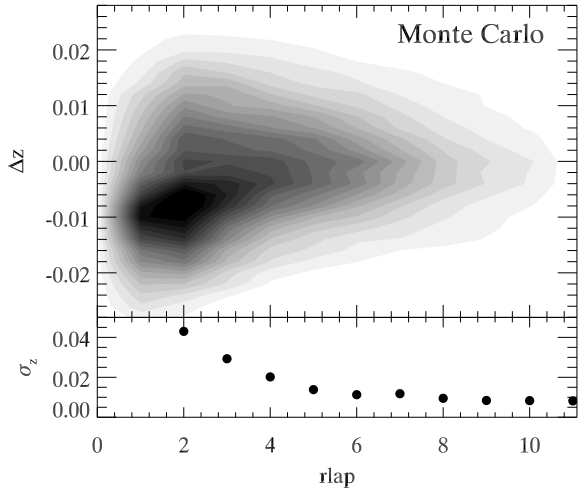}
\caption{{\it Upper panel:} Redshift residuals {\it vs.} $rlap$. {\it
Lower panel:} Standard deviation of redshift residuals in $rlap$ bins
of size unity. For $rlap \gtrsim 5$, $\sigma_z \lesssim 0.01$.
\label{Fig:zsigrlap}}
\end{figure}

We show the distribution of redshift residuals, $\Delta z$ {\it vs.}
the $rlap$ parameter in the upper panel of Fig.~\ref{Fig:zsigrlap},
for input spectra satisfying  $0.3 \le z \le 0.5$; $-5 \le t\ {\rm
[days]} \le +15$ ($t$ is the SN phase); $2 \le$ S/N (per \AA) $\le
10$. The residuals are shown as a two-dimensional histogram, with a
grayscale scheme reflecting the number of points in a given $[\Delta
z,rlap]$ 2D bin (darker for more points). We only show correlations
for which the overlap between input and template spectra $lap \ge
0.40$. For good correlations ($rlap \gtrsim 5$), the distribution of
redshift residuals is a Gaussian centered at $\Delta z = 0$. In the
lower panel, we show the standard deviation of redshift residuals,
$\sigma_z$, in $rlap$ bins of size unity. For $rlap \gtrsim 5$, we
have a typical error in redshift of order $\sigma_z \lesssim 0.01$. 

For poor correlations ($rlap \lesssim 3$) there is a
concentration of points around $\Delta z \approx -0.01$. This is due
to the SN~Ia template phase distribution in our database: many input
spectra at post-maximum phases ($t \lesssim +10$ days) are attracted
to higher phases ($\gtrsim +10$ days), where the position of SN
spectral features has shifted redward in wavelength due to the
expansion of the supernova envelope. The template needs to be shifted
less in $\ln \lambda$ space to match the redshift of the input
spectrum, which leads to an under-estimation of the redshift (by $\sim
0.01$) corresponding to a combination of the typical velocity shift in
SN~Ia absorption features from maximum to $\sim 10$ days past maximum,
and the spread of these velocities at a given phase for different
supernovae ($\sim 3000$\kms; see
\citep{Benetti/etal:2005,Blondin/etal:2006}).

\begin{figure}
\includegraphics[height=.42\textwidth]{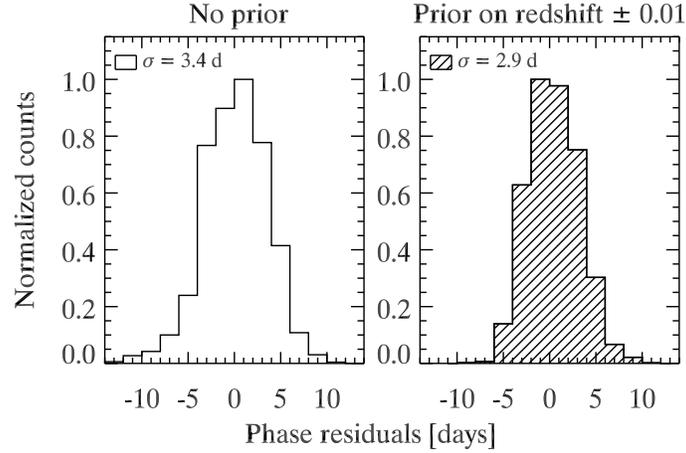}
\caption{Effect of a redshift prior on the phase determination. With
no redshift prior ({\it right}), $\sigma_t = 3.4$ days. With a flat
$\pm 0.01$ redshift prior, $\sigma_t = 2.9$ days.
\label{Fig:deltathist_prior}} 
\end{figure}

This covariance between redshift and phase (over-estimating the
phase leads to under-estimating the redshift, and {\it vice
versa}) suggests that priors on one parameter should improve the
accuracy of the other. Fig,~\ref{Fig:deltathist_prior} shows the
effect on the distributions phase residuals (for $rlap \ge 5$) of
adding a flat $\pm 0.01$ prior on redshift. As expected, a prior on
redshift slightly improves the phase determination ($\sigma_t = 3.4$
days to $\sigma_t = 2.9$ days). A flat $\pm 3$-day prior also improves
the redshift determination ($\sigma_z = 0.006$ to $\sigma_z = 0.004$,
not shown here). In practice, the prior on redshift generally comes
from a spectrum of the SN host galaxy, and one can impose a prior on
phase if a well-sampled lightcurve of the supernova (i.e. one  for
which the date of maximum light is easily determined) is available.

\begin{figure}
\includegraphics[height=.5\textwidth]{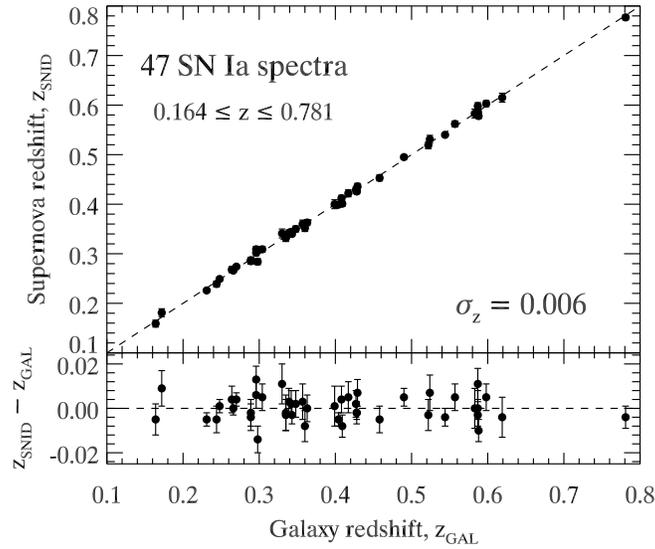}
\caption{{\it Upper panel:} Comparison of redshifts determined from
cross-correlations with SN~Ia templates using SNID ($z_{\rm SNID}$)
and from narrow lines in the host galaxy spectrum ($z_{\rm GAL}$). The
dispersion about the one-to-one correspondence is $\sigma_z \approx
0.006$ over the redshift range $0.164 \le z \le 0.781$. {\it Lower
panel:} Redshift residuals {\it vs.} $z_{\rm GAL}$. The data are from
the ESSENCE project \citep{Matheson/etal:2005,Foley/etal:2007}.
\label{Fig:zgalsnid}}
\end{figure}

We test the accuracy of correlation redshifts using SNID by comparing
it with that obtained from narrow emission/absorption lines in the
host galaxy spectrum. We have selected high-redshift SN~Ia spectra
taken by members of the ESSENCE team
\citep{Matheson/etal:2005,Foley/etal:2007}, for which a redshift of
the host galaxy could be obtained. This amounts to 47 SN~Ia spectra in
the redshift range $0.164 \le z \le 0.781$. The result of this
comparison is shown in Fig.~\ref{Fig:zgalsnid}. The
upper panel is a plot of the supernova redshift determined {\it via}
cross-correlation using \snid, {\it vs.} that determined from narrow
lines in the host galaxy spectrum. The dispersion about the one-to-one
correspondence of the redshifts is excellent, with $\sigma \approx
0.006$ over the whole redshift range. This is in good agreement
with the expected redshift residual found from simulations. The lower
panel shows a plot of the redshift residuals as a function of the
galaxy redshift.

\section{Type Determination}

The previous results are only valid if we assume we know the type of
the input supernova spectrum-- in this case a ``normal''
SN~Ia. Although SNID is tuned to determining SN redshifts, we
investigate its potential in assigning a probability to the input
spectrum being of a certain type. We focus on two distinct examples,
particularly relevant to ongoing high-redshift SN~Ia searches: the
distinction between 1991T-like SNe~Ia and other type-Ia supernovae,
and the increasing difficulty to distinguish between type-Ic
supernovae and SNe~Ia at high redshifts.

\begin{figure}
\includegraphics[height=.65\textwidth,angle=90]{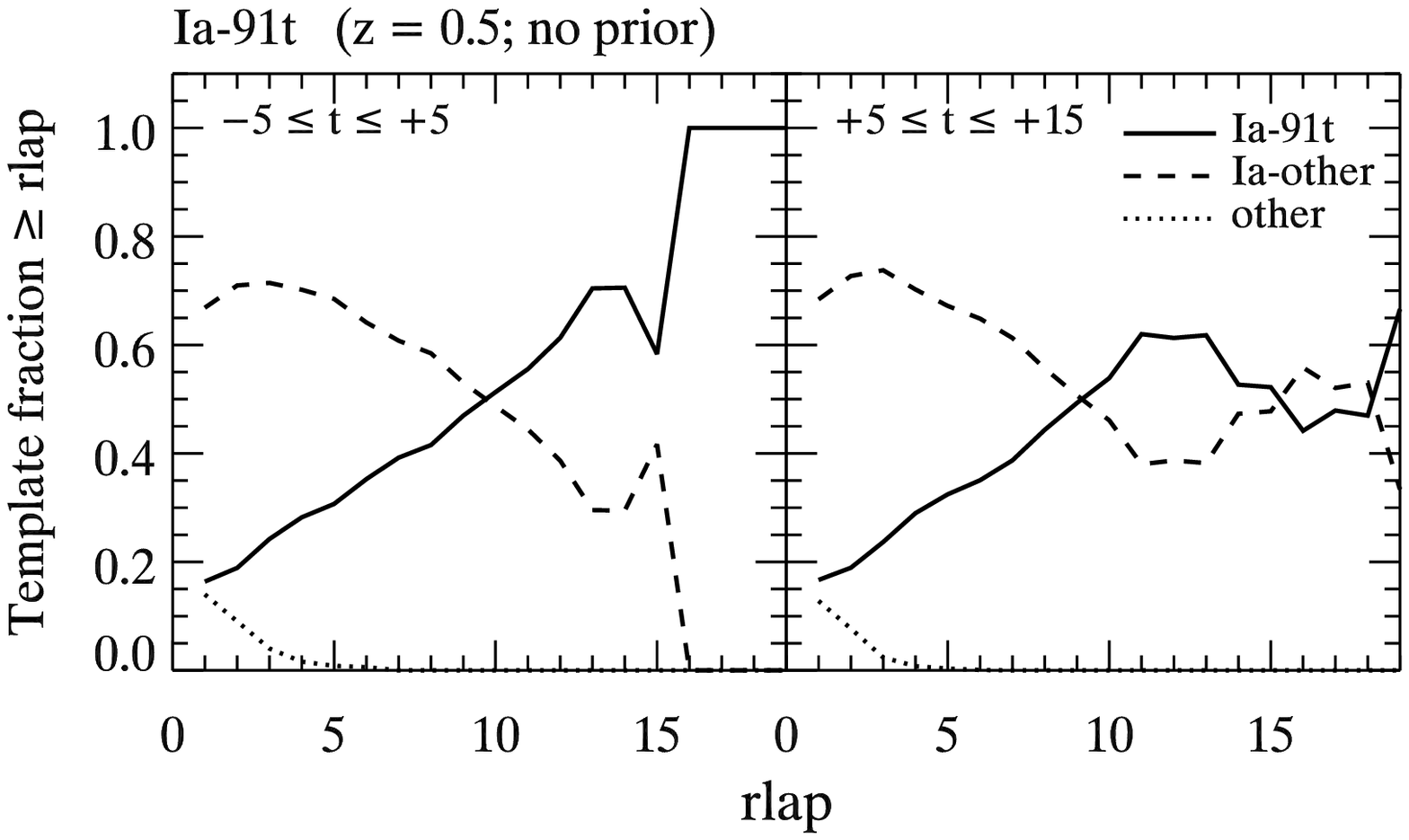}
\hspace*{-1.5cm}
\includegraphics[height=.36\textwidth]{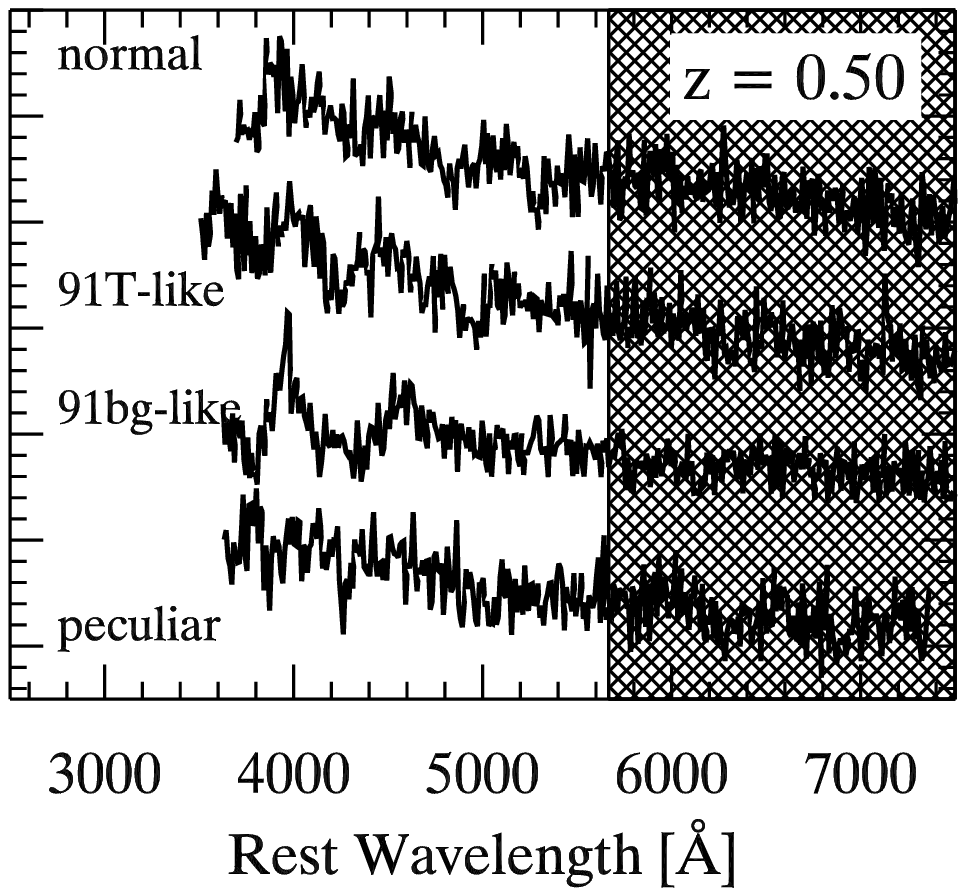}
\caption{Identification of a 1991T-like SN~Ia at $z=0.5$, around
maximum light ({\it right}) and past maximum light ({\it middle}). The
curves correspond to the fraction of templates of a given type greater
than a given $rlap$ value: 1991T-like SNe~Ia ({\it solid}); other
SNe~Ia ({\it dashed}); other SN types ({\it dotted}). The left panel
shows representative maximum-light spectra of the various SN~Ia
subtypes, as observed with a typical optical spectrograph at $z=0.5$.
Note that the relative differences in {\it pseudo}-continuum shapes
has no impact on the SNID results.
\label{Fig:Ia91t}}
\end{figure}

It can be a challenge to distinguish the subtypes of SNe~Ia from one
another (Fig.~\ref{Fig:Ia91t}, {\it right}). 1991T-like SNe~Ia
supernovae have a peak luminosity at the bright end of the SN~Ia
distribution, and although their lightcurves still obey the Phillips
relation \citep{Phillips:1993}, it is useful to have an independent
confirmation of their high intrinsic luminosity {\it via} their
spectra. Spectra of 1991T-like SNe~Ia are characterized by the
near-absence of \catwo\ and \sitwo\ lines in the early-time spectra,
and prominent high-excitation features of \fethree-- not found in
``normal'' SNe~Ia. The \sitwo, \stwo, and \catwo\ features develop
during the post-maximum phases, and by $\sim 2$ weeks past maximum the
spectra of ``1991T-like'' objects are similar to those of ``normal''
SNe~Ia.

In Fig.~\ref{Fig:Ia91t} ({\it left}) we illustrate the ability
for SNID to identify 1991T-like SNe~Ia around maximum light (i.e. when
the spectroscopic differences with ``normal'' SNe~Ia are most
apparent) at $z=0.5$. We show the fraction of templates in the SNID
database that correlate with the input spectrum, as a function of the
$rlap$ parameter: 1991T-like SNe~Ia ({\it solid line}); other SNe~Ia
({\it dashed line}); supernovae of other types ({\it dotted
line}). For $rlap \gtrsim 10$, the fraction of 1991T-like templates
dominates over the other SN~Ia subtypes. The confusion with other SN
types is always low and inexistent for $rlap > 5$. At $\sim 1-2$ weeks
past maximum light, however, the distinction is more difficult to make
(as expected) with roughly a 50\% probability to recover another SN~Ia
subtype (Fig.~\ref{Fig:Ia91t}, {\it middle}).

\begin{figure}
\includegraphics[height=.65\textwidth,angle=90]{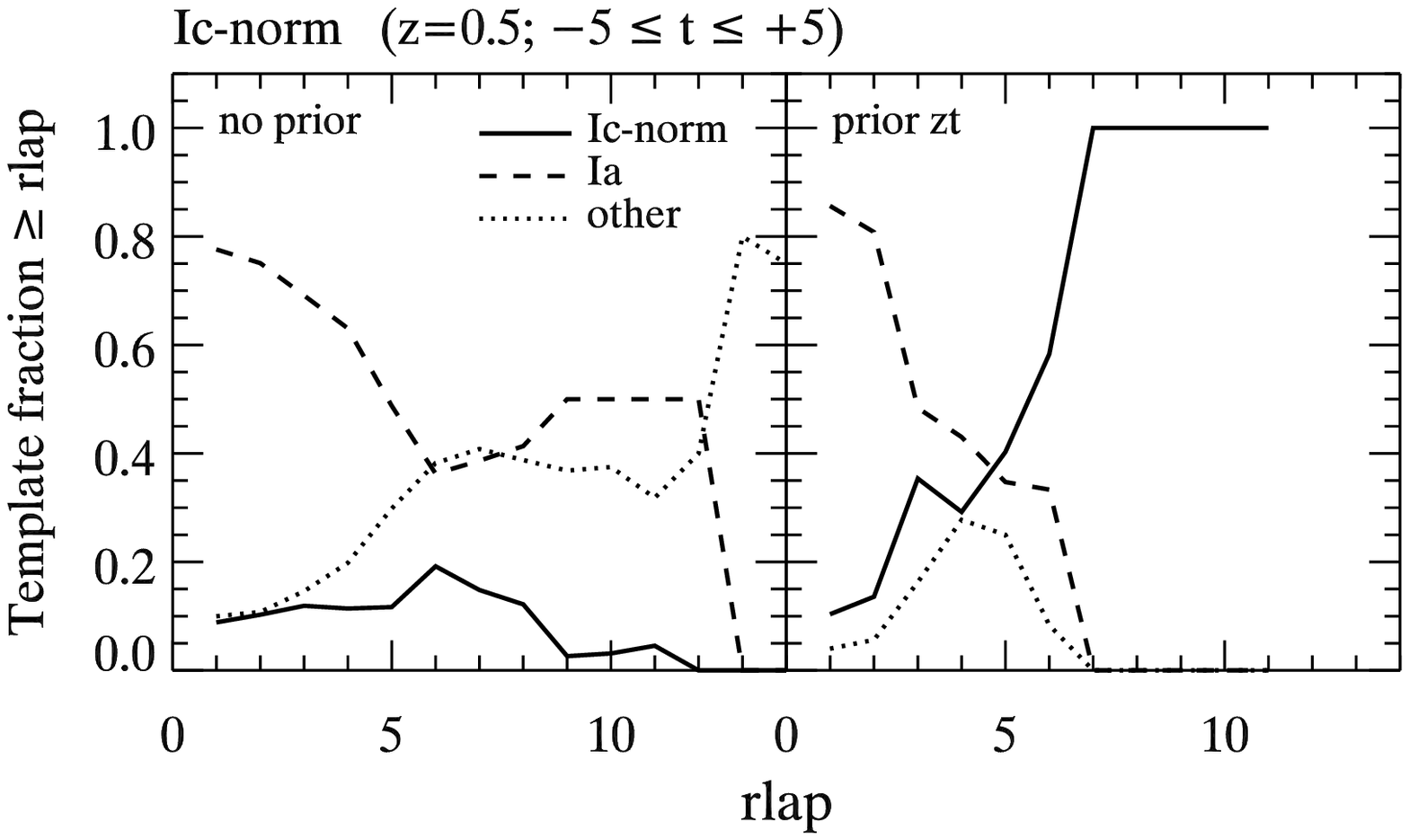}
\hspace*{-1.5cm}
\includegraphics[height=.36\textwidth]{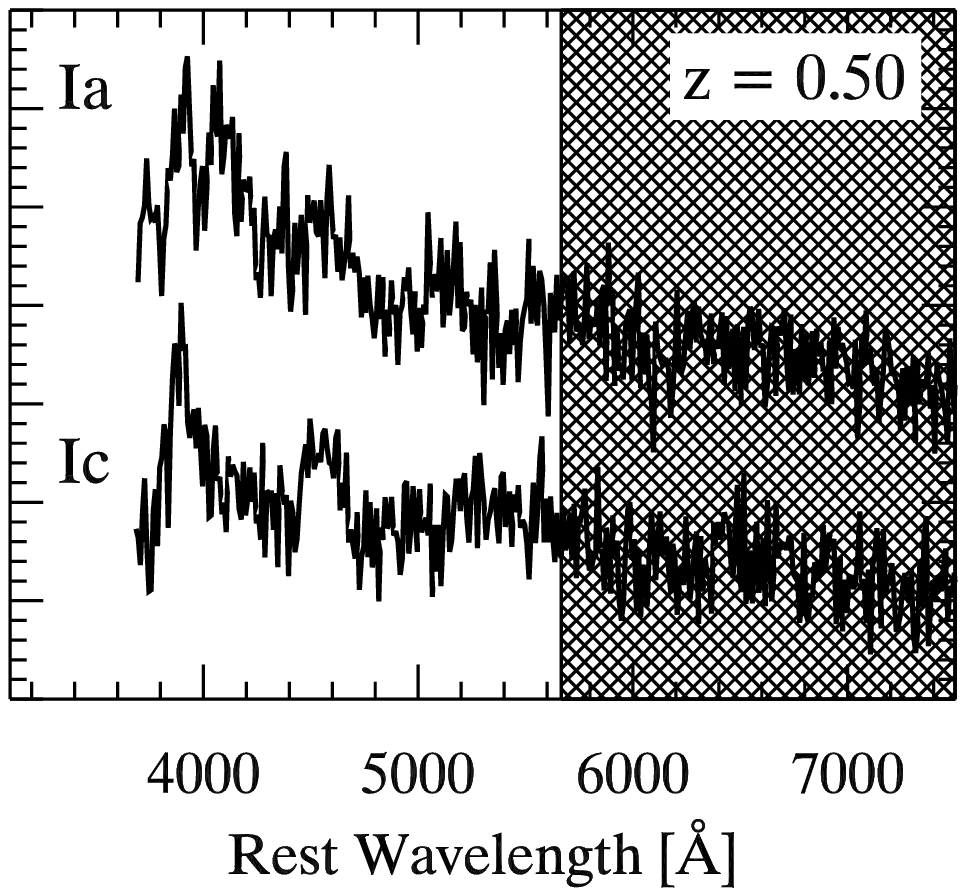}
\caption{Identification of a normal SN~Ic around maximum light at
$z=0.5$, with no prior on redshift or phase ({\it right}), and with a
flat $\pm 0.01$ prior on redshift and a $\pm 3$-day prior on the phase
({\it middle}). The curves correspond to the fraction of templates of
a given type greater than a given $rlap$ value: normal SNe~Ic ({\it
solid}); SNe~Ia ({\it dashed}); other SN types ({\it dotted}). The
left panel shows representative maximum-light spectra of SNe~Ia and
SNe~Ic, as observed with a typical optical spectrograph at $z=0.5$.
Note that the relative differences in {\it pseudo}-continuum shapes
has no impact on the SNID results.
\label{Fig:Ic}}
\end{figure}

The mis-identification of supernova of other types as SNe~Ia is a
major concern for ongoing high-redshift SN~Ia searches
(Fig.~\ref{Fig:Ic}, {\it right}). Including only a small fraction
non-Ia supernovae in a sample will lead to a mis-calibration of the
absolute magnitudes of these objects, and to biases in the derived
cosmological parameters \citep{Homeier:2005}. A particular concern is
the contamination of high-z SN~Ia samples with type-Ic supernovae. At
redshifts $z \gtrsim 0.4$, the defining \sitwo\ $\lambda 6355$
absorption feature of SNe~Ia (also present, though somewhat weaker, in
SNe~Ic) is redshifted out of the range of most optical spectrographs,
and one has to rely on spectral features blueward of this to determine
the supernova type. Some of these features, such as the \catwo\ H\&K
$\lambda \lambda 3934, 3968$ doublet, are common to both SNe~Ia and
SNe~Ic. Other features characteristic of SN~Ia spectra around maximum
light (e.g. \stwo\ $\lambda \lambda 5454, 5640$) are generally weak
and can be difficult to detect in low-S/N spectra. One has to invoke
external constraints, such as the SN color evolution, lightcurve
shape, host galaxy morphology (only SNe~Ia occur in early-type hosts;
\citep{Cappellaro/etal:1997}), or the expected apparent peak
magnitude: type-Ic supernovae at maximum light are often $\gtrsim
1$~mag fainter than SNe~Ia, and hence are not expected to pollute
magnitude-limited samples of SNe~Ia at high redshift (although
\citealt{Clocchiatti/etal:2000} have reported on a type-Ic SN with a
similar absolute magnitude as normal SNe~Ia).

In Fig.~\ref{Fig:Ic} ({\it left}) we illustrate the ability for SNID
to identify SNe~Ic around maximum light ($-5 \le t {\rm \ [days]} \le
+5$) at $z=0.5$. We show the fraction of templates in the SNID
database that correlate with the input spectrum, as a function of the
$rlap$ parameter: ``normal'' SNe~Ic ({\it solid line}); SNe~Ia ({\it
dashed line}); supernovae of other types ({\it dotted line}). For
``good'' correlations ($rlap \ge 5$), the input spectrum almost always
correlates with supernovae of other types. With an additional flat
$\pm 0.01$ prior on redshift and a flat $\pm 3$-day prior on the
phase, the fraction of SN~Ic templates dominates for $rlap \gtrsim 5$
(Fig.~\ref{Fig:Ic}, {\it middle}). In the absence of such priors, one
needs to invoke non-spectral constraints on the type to identify
potential SN~Ic contaminants in high-redshift SN~Ia samples.

\section{Conclusion and Future Work}

We have presented an algorithm, based on the correlation techniques of
Tonry \& Davis \citep{Tonry/Davis:1979}, which can be used to
determine the redshift and phase of a supernova spectrum and place
constraints on its type. We develop a diagnostic, the $rlap$
parameter, to quantify the quality of a given correlation between the
input and a template spectrum. This parameter is simply the
product of the Tonry \& Davis $r$-value and the overlap, $lap$,  in
$\ln \lambda$ space between the input and template spectrum at the 
correlation redshift. We show, based on simulations, that for $rlap
\gtrsim 5$, the typical error on redshift and phase is $\sigma_z \lesssim
0.01$ and $\sigma_t \lesssim 3$ days, respectively. The former
accuracy on redshift is confirmed through a comparison of correlation
redshifts with host-galaxy redshifts (determined from narrow lines in
the spectrum) out to redshifts $z \lesssim 0.8$.

We present first results of an impartial and effective spectroscopic
classification of supernovae, based on the cumulative fraction of
correlations exceeding a certain $rlap$ cutoff. We illustrate this
through two examples, relevant to ongoing SN~Ia searches at high
redshift: we are able to distinguish 1991T-like SNe~Ia from other
SNe~Ia at $z=0.5$, if the input spectrum is within five days from
maximum light; we identify a type-Ic supernova as such at $z=0.5$, but
only when an additional prior on redshift and phase is applied. These
examples both illustrate the success and limitations of such an
automated classification scheme, and highlight the complementarity
between spectroscopic and photometric observations in determining the
supernova type.

The current version of SNID will soon be made available to the
community, and we plan to set up a web-based interface for
instantaneous supernova typing (and redshift/phase
determination). Future versions of SNID will include a
wavelength-weighted $lap$ parameter, an explicit treatment of the
covariance between redshift and phase, and a Bayesian approach to type
determination, as currently used for photometric classification of
supernovae
\citep{Poznanski/Maoz/Gal-Yam:2006,Kuznetsova/Connolly:2006}.
Moreover, more spectral templates are continuously being included in
the SNID database, which directly impact the ability for SNID to
securely identify input spectra.




\begin{theacknowledgments}
This work has been funded in part by the US National Science
Foundation through grants AST 0443378, AST 057475, and
AST 0606772. This research has made use of the CfA Supernova Archive,
which is funded in part by the National Science Foundation through
grant AST 0606772.
\end{theacknowledgments}



\bibliographystyle{aipproc}   

\bibliography{blondin}

\end{document}